\documentclass[sigconf]{acmart}

\usepackage{balance}
\usepackage{upgreek}
\usepackage[ruled,linesnumbered]{algorithm2e}
\SetKwRepeat{Do}{do}{while}%

\usepackage[utf8]{inputenc}
\usepackage[english]{babel}

\usepackage{amsmath}
\usepackage{centernot}

\usepackage{mathtools}

\usepackage{xparse}
\DeclarePairedDelimiterX{\set}[1]{\{}{\}}{\setargs{#1}}
\NewDocumentCommand{\setargs}{>{\SplitArgument{1}{;}}m}
{\setargsaux#1}
\NewDocumentCommand{\setargsaux}{mm}
{\IfNoValueTF{#2}{#1} {#1\,\delimsize|\,\mathopen{}#2}}

\newcommand{\minus}{\scalebox{0.75}[1.0]{$-$}}

\newtheorem{theorem}{Theorem}

\newtheorem{observation}[theorem]{Observation}

\copyrightyear{2020}
\acmYear{2020}
\setcopyright{acmlicensed}
\acmConference[UMAP '20] {28th ACM Conference on User Modeling, Adaptation and Personalization}{July 14--17, 2020}{Genoa, Italy}
\acmBooktitle{28th ACM Conference on User Modeling, Adaptation and Personalization(UMAP'20), July 14--17, 2020, Genoa, Italy}
\acmPrice{15.00}
\acmDOI{10.1145/3340631.3394858}
\acmISBN{978-1-4503-6861-2/20/07}

\begin{document}

%
\title[Using Stable Matching to Optimize the Balance between Accuracy and Diversity]{Using Stable Matching to Optimize the Balance between Accuracy and Diversity in Recommendation}

%

\author{Farzad Eskandanian}
\email{feskanda@depaul.edu}
\affiliation{%
  \institution{DePaul University}
  \city{Chicago}
  \state{Illinois}
  \country{USA}
}

\author{Bamshad Mobasher}
\email{mobasher@cs.depaul.edu}
\affiliation{%
 \institution{DePaul University}
 \city{Chicago}
 \state{Illinois}
 \country{USA}}

\begin{abstract}
Increasing aggregate diversity (or catalog coverage) is an important system-level objective in many recommendation domains where it may be desirable to mitigate the popularity bias and to improve the coverage of long-tail items in recommendations given to users. This is especially important in multistakeholder recommendation scenarios where it may be important to optimize utilities not just for the end user, but also for other stakeholders such as item sellers or producers who desire a fair representation of their items across recommendation lists produced by the system. Unfortunately, attempts to increase aggregate diversity often result in lower recommendation accuracy for end users. Thus, addressing this problem requires an approach that can effectively manage the trade-offs between accuracy and aggregate diversity. In this work, we propose a two-sided post-processing approach in which both user and item utilities are considered. Our goal is to maximize aggregate diversity while minimizing loss in recommendation accuracy. Our solution is a generalization of the Deferred Acceptance algorithm which was proposed as an efficient algorithm to solve the well-known stable matching problem. We prove that our algorithm results in a unique user-optimal stable match between items and users. Using three recommendation datasets, we empirically demonstrate the effectiveness of our approach in comparison to several baselines. In particular, our results show that the proposed solution is quite effective in increasing aggregate diversity and item-side utility while optimizing recommendation accuracy for end users.
\end{abstract}

\maketitle

\section{Introduction}

Much of the early research in recommender systems has been focused on improving the accuracy of personalized recommendations. These efforts have led to a number of effective recommendation algorithms and strategies. In recent years, however, more attention has been focused on key objectives other than accuracy that may also have a bearing on the overall satisfaction of users as well as other possible stakeholders of the underlying system. One of these key objectives is recommendation ``diversity''. 

Broadly speaking, two notions of diversity have been explored in the context of recommendation: {\em individual diversity} and {\em aggregate diversity}. Individual diversity reflects the ability of the recommender systems to recommend a broad range of items to individual users (for example, items ranging over multiple categories, or spanning both popular items as well as niche items). This ability is often measured as a function of the dissimilarity of items in each recommendation list provided to users. Improving individual recommendation diversity is an important objective since it can lead to higher user satisfaction and helps reduce the ``filter bubble'' effect \cite{nguyen2014FilterBubble} affecting many personalized systems. 

Aggregate diversity, on the other hand, measures the ability of the system as a whole to recommend a wide selection of items. This is an important system-level objective that ensures a fair opportunity for most items to be recommended. Increasing aggregate diversity helps mitigate the popularity bias by increasing the coverage of niche or less popular items. It also promotes fairness from the perspective of other stake-holders such as sellers or suppliers who need to have a guarantee that the system provides a fair coverage of their items across recommendation lists. Aggregate diversity is often measured as a function of the number or the ratio of items that are recommended to users relative to the set of all items available across all suppliers. 

In this paper we focus primarily on aggregate diversity especially as an objective related to item side of the recommendation scenario. In particular we are interested in a two-sided framework that can produce a balanced match between items and users taking into account both user utilities (reflected in recommendation accuracy) as well as item utilities (reflected in aggregate diversity).

As with other beyond-accuracy objectives, there is often a trade-off between aggregate diversity and personalization or recommendation accuracy: approaches that try to increase aggregate diversity generally do so at the cost of accuracy degradation. A variety of approaches have been proposed to manage this trade-off including optimization approaches using network flow models on the bipartite user-item graph \cite{adomavicius2011maximizing, adomavicius2012improving, antikacioglu2017post} as well as re-ranking methods involving utilizing the dual recommendation perspective of recommending (matching) items to appropriate users \cite{vargas2014salesdiv}. These approaches generally try to balance the diversity-accuracy trade-offs either by trying to optimize for recommendation accuracy while imposing certain constraints on the items, or by using multi-criteria optimization where both accuracy and diversity are optimized individually and the trade-off is controlled through appropriate parameter settings. 

We introduce a more principled two-sided approach to managing the diversity-accuracy trade-off that takes into account both user and item utilities. User utilities are maximized when the ranking accuracy of the recommendation lists are maximized. Item utilities, on the other hand, are maximized when the item is matched with the most suitable list of users. We view the problem as an assignment problem where the goal is to distribute items to each user’s recommendation list of size $k$, such that both user and item utilities are maximized and where each item is subject to a capacity constraint $C$ (i.e. can be recommended at most $C$ times). We propose a generalization of a well-known and efficient algorithm for solving the two-sided {\em stable matching} problem adapted to recommendation context. Specifically, we extend the {\em Deferred Acceptance} (also known as Gale-Shapely) algorithm \cite{gale1962college} used to solve the college admission problem. In this problem, given a set of students and set of colleges with quotas the goal is to find the best (stable) matching possible based on the ranking preferences of the two groups. In our case, the ranking preferences for the user-side and the item-side are specified based on user and item utility functions, respectively. The algorithm then automatically finds the best stable match possible subject to item cap constraints. 

We show that the proposed algorithm is user-optimal (maximizes user utilities) subject to the cap constraint. But the algorithm also attempts to maximize item utilities while maintaining a stable match (when there is a symmetry between user and item preferences), and this, in turn, results in higher overall aggregate diversity. From a practical perspective, our proposed algorithm provides added flexibility as a post-processing approach because it can be used in conjunction with any top-$N$ recommendation algorithm that can generate preference rankings for users and items. In our experiments, we empirically demonstrate the effectiveness of our approach in comparison to several baselines and using different underlying recommendation algorithms. Our results show that our approach is effective in increasing aggregate diversity and item-side utility while maintaining or improving recommendation accuracy for end users.

\section{Related Work}

As noted earlier, improving individual recommendation diversity has been shown to lead to better user satisfaction \cite{pu2011userStudyDiv, jones2007userStudyDiv} and to reduce the ``filter bubble'' effect \cite{nguyen2014FilterBubble} affecting many personalized systems. It is also an important consideration for reducing popularity bias \cite{park2008long, yin2012challengingLong} that affects many collaborative recommender systems. There has been an extensive research in developing methods for increasing individual diversity and in understanding the trade-offs between diversity and recommendation accuracy \cite{zhang2008MonotonyDiv, ziegler2005improving, eskandanian2017clusteringDiv_evaluating, eskandanian2016Diversity, vargas2011rankRelDiv, puthiya2016coverage}. Other related notions affecting the quality of recommendations to end users, such as novelty and serendipity, have also been extensively explored \cite{adamopoulos2015unexpectedness, Kaminskas:2016:DSN:3028254.2926720}.

Increasing aggregate diversity, on the other hand helps mitigate the popularity bias by increasing the coverage of items recommended from the long-tail \cite{celma2008hitsPopBias, yin2012challengingLong, park2008long, anderson2006long}. In prior work, aggregate diversity has been referred to using different terms including as “sales diversity” \cite{fleder2009blockbusterSalesDiv, vargas2014salesdiv, kelly2006enhancing, jannach2013recommenders}, ``catalog coverage'', or simply “coverage” \cite{Kaminskas:2016:DSN:3028254.2926720}. In this paper we use the term ``aggregate diversity'' to broadly encompass these other representations of the concept.

In \cite{graphReRanking2017post, adomavicius2011maximizing} a graph-theoretic approach has been used to increase aggregate diversity and also maintaining accuracy. Specifically, \cite{graphReRanking2017post} formulated the recommender system as a subgraph selection problem from a supergraph generated by any recommender system in a bipartite graph of users and items. Then, they solved this problem using a Minimum-cost network flow approach. In their approach a similar constraint has been used to limit the number of times each item can be recommended. Their goal is not to strictly satisfy this constraint but to penalize the recommendations that try to deviate from this constraint. Our approach and methodology is completely different from theirs. First, the mentioned constraint will be always satisfied completely in the resulting re-ranked recommendations. Secondly, the stable matching proposed in this paper is a simple and efficient (polynomial time complexity) method both in terms of implementation and ease of understanding. Besides that, our problem is a two-sided recommendation where utilities of both users and items are incorporated into the model. However, in their approach only user-side utilities are considered.

Another line of works which has been proposed recently by \cite{burke2016MSP} is to consider the objectives of different multistakeholders in the system. For example, in any recommender system three stakeholders can be defined: consumers, providers and the platform for these two to be connected. Good examples of such platforms are job recommendation and item recommendation in Amazon. In each of these platforms the consumers and providers are separated from the platform and the utility of both parties are important \cite{nasim2018balanced}. Very recently \cite{ozge2018multistakeholder} has formulated this problem using an integer linear programming approach. Their goal is to maximize utility of both consumers and providers using an optimization method. Our approach is similar in terms of overall goals. However, in our problem we incorporate the utility of items in to the optimization while they only consider the item utility as a constraint on the optimization that needs to be satisfied. In other words, they do not take into account the ranking preferences of items over users as we do it in our method. In addition to that, our algorithm is completely different and draws on well-established Gale-Shapely algorithm \cite{gale1962college} for the stable matching problem. The proposed method in this paper is computationally more effective than the integer programming approach.


\section{Background and Problem Definition}
\label{sec:definitions}

In a typical recommender system there are two types of entities: Users and Items. Personalization is the main feature of recommender systems allowing the system to match users with items that are consistent with those users' preferences. A recommender system can be viewed as a bipartite graph $G = (\mathcal{U}, \mathcal{I}, \mu)$, where $\mathcal{U}$ is the set of all users, $\mathcal{I}$ is the set of all items and $\mu$ is the set of edges represented as user-items pairs $(u,i)$. The goal of a recommender system is to rank the items such that preferred items by each user appear higher in the ranking. The top $N$ items in the ranking are then presented to the user as recommendations. 
For our purposes, we do not make any assumptions about which underlying algorithm is used to rank the items, but assume that such a ranking can be produced. We denote the preference ordering of user $u$ on all items in $I$ based on the recommender system ranking by $\succ_u$. We say user $u$ prefers items $i_1$ to $i_2$ if $i_1\succ_u i_2$. 

The preference ordering can be used to define the notion of {\em user utility}. As we shall see in the following, we use Normalized Discounted Cumulative Gain (NDCG) to measure user utilities, however, other ranking measures can also be used for this purpose.

In the typical recommendation setting, the goal is usually to maximize user utilities.
However, a single-minded focus on user utilities without considering the utility of items may result in an unfair or uneven distribution of items among all the recommendation lists. In reality, there are often few popular items that receive the lion share of user attention in the system. Especially in collaborative recommender systems, the abundance of user feedback on popular items results in recommendations that are biased towards these items, perpetuating the popularity bias. Thus, one-sided recommender systems (that are focused solely on maximizing user utilities, tend to produce a power law distribution of recommended items and an unfair platform for most of the items to compete with a few popular items. This, in turn, may reduce the incentive to participate on the part of many sellers or producers.  

Ideally, the distribution of recommended items should be uniform with maximum entropy. However, since the user preferences are concentrated over a small set of items, any attempts to ``flatten'' the distribution of items in recommendations will result in significant loss in accuracy and personalization effectiveness. Ideally, we would like to maximize the entropy of item distribution across recommendations while minimizing accuracy loss. 

One way to control the entropy of item distribution is to impose a constraint $C$ over the capacity of items (the number of times each item can be recommended $n_i$). This is the approach taken in several prior methods that aim to increase aggregate diversity while optimizing user utilities \cite{graphReRanking2017post}. This approach,however, ignores potentially relevant information available on the item side of the recommendation platform. Just as user preference rankings of items can be used as the basis for user-side utility function to provide the most relevant items to users, a dual notion of item utility can be used to represent the best match between each item and a ranked list of users. Indeed, increasing aggregate diversity in a brute-force fashion does not necessarily mean that each item is matched with the most appropriate set of users. 

Our goal in this work is to automatically achieve a balance between accuracy and aggregate diversity that takes into account both user and item utilities.  Specifically, our approach re-ranks the recommendations for users by maximizing both user and item utilities subject to a capacity constraint $n_i \leq C$ in such a way to create a stable match between users and items. We do so by extending the well-known Gale-Shapely or {\em Deferred Acceptance} algorithm \cite{gale1962college} in order to handle the many-to-many matching problem which is required in the context of recommender systems. In the following we provide a more formal statement of the problem and in the next section we present the details of our algorithm and discuss some of its properties. \\

\noindent\textbf{User Utility:} \\

We use Normalized Discounted Cumulative Gain (NDCG) as the basis for defining our notion of utility. NDCG is a well-known measure of ranking quality in information retrieval which measures the gain of each item based on the change in the ranking position of the item in a ranked list of items. Given the ranking preference of a user $\succ_u$ and her recommendation list $L$, the Discounted Cumulative Gain (DCG) is defined as:
\begin{equation} \label{dcg_eq}
    DCG(L, \hspace{1.5pt} \succ_u) = \sum_{i \in L}{\frac{rel(i, \ \succ_u)}{log_2(i+1)}}
\end{equation}
Relevance scores can be defined in a small range of values such as $rel=(0, r_{max})$. The relevance score of each item for user $u$ is defined based on its ranking position in $\succ_u$:
\begin{equation}
    rel(i, \hspace{1.5pt} \succ_u) = \frac{rank^{-1}(i, \ \succ_u)}{r_{max}}
\end{equation}
where the numerator is the reversed ranking position of item $i$ in user $u$'s ranking preference $\succ_u$. 

Given a user ranking preference order $\succ_u$ and any other item ranking $L$, the gain in user $u$'s utility relative to $L$ is defined as:
\begin{equation} \label{ndcg_user_util}
    NDCG_u(L) = \frac{DCG(L, \ \succ_u)}{IDCG(u)}
\end{equation}
where $IDCG(u)$ is the ideal $DCG$ of user $u$ in which the relevance scores for all items in equation \ref{dcg_eq} is equal to the maximum value $rel(i)=r_{max}$. 

In a one-sided recommendation setting the problem can be formulated as a user utility optimization problem subject to certain constraints. Specifically, for each user $u$ a recommendation list is a set of user-item pairs $\mu_u$. The user-side utility of the matching $\mu_u$ is computed using $NDCG_u(\mu_u)$. Therefore, given the item capacity $C$ and set of all the matchings (recommendations) $\mu = \{\bigcup_{u \in \mathcal{U}} \mu_u\}$ the problem formulation would be as follows:
\begin{equation} \label{ConstraintUserUtilMaxmization_eq}
    \begin{aligned}
    & \underset{\mu}{\text{maximize}}
    & & Utility_\mathcal{U}(\mu) \\
    & \text{subject to}
    & & n_i \leq C, \; i \in \mathcal{I}.
    \end{aligned}
\end{equation}
where $Utility_\mathcal{U}(\mu) = NDCG_\mathcal{U}(\mu)$ and $n_i$ is the number of times item $i$ is recommended.

As noted earlier, however, our goal is to optimize both user and item utilities. \\

\noindent\textbf{Item Utility:} \\

Similar to user-side utility, the item utility is defined by an item-side version of $NDCG$. Given an item $i$ ranking preference $\succ_i$ over all the users, the gain in item $i$'s utility given a matching $\mu_i = \{(u,i), (u', i), ...\}$ is defined as follows:
\begin{equation}
    Utility_{\mathcal{I}}(\mu_i) = NDCG_i(\mu_i) = \frac{DCG(\mu_i, \ \succ_i)}{IDCG(i)}
\end{equation}
Similar to equation \ref{ndcg_user_util}, the ideal $DCG$ for item $i$ is denoted by $IDCG(i)$. 

In a two-sided recommendation the problem can be defined as simultaneously optimizing utilities of users and items:
\begin{equation}
    \begin{aligned}
    & \underset{\mu}{\text{maximize}}
    & & \Big(Utility_\mathcal{U}(\mu), \ Utility_\mathcal{I}(\mu)\Big) \\
    & \text{subject to}
    & & n_i \leq C, \; i \in \mathcal{I}.
    \end{aligned}
\end{equation}

It is important to note that there is an asymmetry between ranking preferences of items and those of users. This means that utilities of each side is derived independently and are not necessarily consistent with each other. For example, if user $u$ prefers item $i$ the most, it does not always hold that in item $i$'s preference ordering, $u$ is the highest ranked user, i.e.,
\begin{equation}
    \{\forall i' \in I : (i \succ_u i')\} \centernot\iff  \{\forall u' \in U : (u \succ_i u')\}  
\end{equation}
Therefore, finding a matching that maximizes the utility of one side does not necessarily maximize the utility of the other side. Popularity bias increases this asymmetry and this makes the simultaneous optimization of both sides very difficult requiring an intelligent way to manage the trade-off point between the two utilities.

\section{Stable Matching and Deferred Acceptance Algorithm}

In the previous section we formulated the assignment problem such that the utilities of both user and item sides are considered. In this section, we introduce an efficient solution to this problem and show the correctness of the algorithm. 

The Deferred Acceptance (also known as Gale-Shapely) algorithm \cite{gale1962college} is a well-known method for solving the college admission problem. In this problem, given a set of students and set of colleges with quotas the goal is to find the best matching possible based on the ranking preferences of the two groups. Gale-Shapely's method is based on the key notion of a ``stable match''.

In the recommendation context, a matching $\mu$ is said to be unstable if there exists another matching $\mu'$ where both utilities of users and items have increased in $\mu'$ compared to $\mu$. In other words, if $i \succ_u i'$ and $u \succ_i u'$ and also $\{(u,i'), (u',i)\} \in \mu$, then $\mu$ is unstable because both $u$ and $i$ can be better matches for each other and therefore both user and item utilities are higher in $\mu'$. A matching $\mu$ is stable if no such unstable match exist for every pair of matches $\{(u,i), (u',i')\}$ $\in$ $\mu$. Gale and Shapely showed that a stable match always exists and the Deferred Acceptance algorithm always finds a stable match. The output of this algorithm is a one-to-many matching where each college will be matched to many students based on their quotas. The ingenuity of the Gale-Shapely's method is in this intuitive strategy of solving a combinatorial optimization problem.

\begin{algorithm}[t]
\caption{MMDA: Many-to-Many Deferred Acceptance}
\label{matching_algo}
\KwIn{The ranking preference of users and items: $\{\succ_u$ and $\succ_i$: $\forall u \in \mathcal{U}$ and $i \in \mathcal{I}\}$; $k$: recommendation list size for each user; $c_i$: item caps.}
\KwOut{User-Item Matching $\mu$}
    $\mu \gets \emptyset$
    
    \For {$u \in \mathcal{U}$}{
        $S_u \gets \emptyset$
    }
    
    \While {$ \set*{\exists u  \hspace{0.3em}; \hspace{0.4em}|\mu_u| < k}$}{
        $i \gets Top \left( \succ_u \ \notin \ S_u \right)$
        
        $\mu_u \gets \mu_u \cup i$
        
        $S_u \gets S_u \cup i$
    
        \If {$|\mu_i| > c_i$} {
        
        $\Phi_i \gets \text{sort}^{-1} (\mu_i) \hspace{0.9em} \text{subject to} \ \succ_i$.
        
        $\mu_i \gets \mu_i \setminus \set*{ \forall u \in \mu_i \hspace{0.3em}; \hspace{0.3em} \text{rank}(u \in  \Phi_i) > c_i }$ 
        }
    }
    {return $\mu$}
\end{algorithm}

The input to the Deferred Acceptance Algorithm is a fully-connected directed bipartite graph connecting users and items. There are directed edges in each direction between user and item nodes. There is a preference ranking for each node on each side of this graph. The algorithm begins to match user-item pairs starting from one side of the bipartite graph (here we assume it starts from the user side). A user $u$ is matched with an item $i$ in increasing order of her rankings. At a given rank $k$, two situations are possible. First, item $i_k$ ($k$th item in user's ranking) may not be currently matched with any user. In this case, a match $(u , i_k)$ is created. On the other hand, item $i_k$ may already be part of a match with another user $v$. In this case if $i_k$ prefers $v$ over $u$, then user $u$’s next ranked item is considered since the current match $(v , i_k)$ is more stable. Otherwise, it is replaced with the more stable match $(u , i_k)$ and user $v$ is considered again for subsequent matches. This process is repeated until all users are matched with items. This would be the one-to-one matching of Deferred Acceptance algorithm. 

We extend this algorithm by allowing the members of each side to have more than one matching. This way each user can be matched to many items and hence receive these items as recommendations. We introduce the $cap$ for items as a constraint to limit the number of times an item can be matched (recommended) to users. Therefore, the combinatorial challenge of this problem would be to find the best matching that cannot be improved further (no other matching exist that would be better off for both sides). We call this algorithm the Many-to-Many Deferred Acceptance (MMDA) algorithm (see Algorithm~\ref{matching_algo}). Next, we introduce the notation that is required for this algorithm. Two functions are used to work with preference rankings in the algorithm. The first function is $Top \left( \succ_u \ \notin \ S_u \right)$ which returns the top-ranked item $i$ from $\succ_u$ where $i$ does not belong to $S_u$. The second function is $[\text{sort}^{-1} (\mu_i) \ ,\text{subject to} \ \succ_i]$ which reverse sort (or rank) the items in $\mu_i$ based on ranking preferences in $\succ_i$.

The algorithm starts with $\mu$ set to empty set. Also, for each user we initialize an empty set to $S_u$ for the set of selected items by $u$. As we iterate over users' preference rankings $\succ_u$ in the main loop of the algorithm (line 5), we add items to $S_u$. The main while-loop continues until we assign $k$ items to each user's matching set $\mu_u$. This makes sure that we have recommended at least $k$ item to each user. Next at line 6, we set item $i$ as the top preferred item in user $u$'s ranking which has not yet been selected. At line 7 we assign item $i$ to the user $u$'s matching and also, adding $i$ to the selected items of this user $S_u$. At line 9, we check to see whether any of the item constraints are violated. If so, we should find the ranked list (subject to $\succ_i$) of assigned users to item $i$ and, reject the least preferred user for item $i$ by removing it from the assigned users $\mu_i$ to item $i$. Next, we discuss the observations and theorems that are followed by MMDA.

\noindent\begin{observation} \label{Obs_itemsGetBetter}
Once item $i$ is assigned to a user $u$, in later iterations of the algorithm either the assignment $(u,i)$ remains the same or it gets better $(u',i)$ where $u' \succ_i u$. In other words, the utility of each item's assignment is monotonically non-decreasing.
\end{observation}

\noindent\begin{observation} \label{Obs_usersGetWorse}
On the other hand, individual users' utilities get worse over later iterations of the algorithm. Users always start with their best options and in case of rejections they look for less preferred items in their ranking preference.
\end{observation}

\noindent\begin{theorem} \label{stableAlgo_theorem}
MMDA always finds a stable matching. 
\end{theorem}

\begin{proof}
(by contradiction). Let's assume there is an assignment which makes $\mu$ from algorithm \ref{matching_algo} unstable. Therefore, by definition there should be an unstable assignment for users $u$ and $u'$ and items $i$ and $i'$, where all of the followings are true:
$u \succ_i u'$, $u \succ_i u'$, $(u, i) \not\in \mu$, $(u, i') \in \mu$, and $(u', i) \in \mu$.

Since $u$ prefers $i$, according to the observation \ref{Obs_usersGetWorse} the algorithm should assign $i$ before assigning $i'$ to $u$. Now, there can be two possible cases for $u'$:

    \noindent\textbf{Case 1:} $u'$ is assigned to $i$ in a previous iteration. At this point, $u$ first selects $i$. Now, $i$ should decide between $u'$ and $u$. Based on observation \ref{Obs_itemsGetBetter} items always choose the better user according to their ranking preferences. Therefore, $\mu$ rejects the previous assignment $(u',i)$ and accept the new assignment $(u,i)$. This contradicts the assumption that $\{(u,i'), (u',i)\} \in \mu$. 
    
    \noindent\textbf{Case 2:} $u'$ is assigned to $i$ after $u$ selecting item $i$ and matched with it in $(u,i)$. Since, by definition $i$ prefers $u$ to $u'$, then $i$ will remain matched with $u$ contradicting the instability assumption.
\end{proof}

\begin{theorem} \label{user_optimal_theorem}
MMDA is user-optimal. 
\end{theorem}

\begin{proof}
(by contradiction). Let's assume that the output $\mu$ of algorithm \ref{matching_algo} is not user-optimal. Therefore, there should exist another matching $\mu'$ where at least a pair of assignments is different so that $Utility_U(\mu') > Utility_U(\mu)$. We can represent the difference between $\mu$ and $\mu'$ is as follows:
\begin{equation*}
    \mu = \{(u,i), (u',i')\} \cup \{\mu \cap \mu'\}
\end{equation*}
\begin{equation*}
    \mu' = \{(u,i'), (u',i)\} \cup \{\mu \cap \mu'\}
\end{equation*}
In the following, we show that if $\mu'$ exists with a greater user utility satisfying the constraint $n_i\leq C$, then there is a contradiction. To show this we consider two cases: 

\noindent\textbf{Case 1:} ($i \succ_u i'$). If user $u$ prefers item $i$ more than $i'$, then the user-side utility of $\mu$ is greater than $\mu'$ and therefore $\mu$ is more user-optimal than $\mu'$.

\noindent\textbf{Case 2:} ($i' \succ_u i$). In this, case assuming that there is a competition between $u$ and $u'$, user $u'$ should also be more interested in $i'$. There can be two cases with respect to the preferences of item $i'$: either $u \succ_{i'} u'$ or $u' \succ_{i'} u$. Again we break down these two cases.

    \noindent\textbf{Case 2.1:} Our assumptions are: $i' \succ_u i$, $i' \succ_{u'} i$, $u \succ_{i'} u'$.
Based on these assumptions for this case $i'$ prefers $u$ over $u'$ and also $u$ prefers $i'$ over $i$. Then by definition the existence of assignment $(u,i) \in \mu$, makes $\mu$ unstable. Which contradicts with theorem \ref{stableAlgo_theorem}.
    
    \noindent\textbf{Case 2.2:} $(u' \succ_{i'} u)$. Since, $i' \succ_{u'} i$ is true as an assumption, then both user and item utilities are greater for $\mu$ than $\mu'$.
\end{proof}

Note that there may be many stable matchings but there is only one user-optimal stable matching (given that we assume users make the first selection of items in their ranking preferences over items). Also, in the situation where too many users (more than the cap) are interested in one item users who are not preferred by that item will be rejected in order to maintain the stability of the match. In the experiments we show how aggregated user and item utilities increase with each given iteration of the algorithm. The monotonicity of the utilities provides empirical evidence of the correctness of this algorithm.


The computational complexity of the MMDA is $O(n\times m \times k \ log \ k)$ where $n$ and $m$ are the number of users and items in the system. The worst-case scenario is when every user should go over all of his/her ranking preferences of $m$ items at most and each time sort the items in $\mu_i$ where $|\mu_i| = k$ is the recommendation list size.

Controlling utility of each side can be done using item caps. MMDA strictly tries to satisfy the items' capacity constraint which is the only parameter of this algorithm. However, there is a limit to the increase of item-side utility if the algorithm is user-optimal. To go beyond this limitation the item-optimal version of this algorithm can be used where items start selecting users first based on their preferences.

\section{Experiments}

In the previous section we proved that the MMDA algorithm is both stable and user-optimal. In this section we provide further empirical evidence for the effectiveness of the algorithm using three different data sets and under different conditions. In our experiments the resulting recommendations are evaluated for each side in isolation. On the user side, we use traditional evaluation metrics for accuracy and individual diversity of recommendations, while on the item-side we measure changes in item utilities. We also use standard metrics such as the Gini-index to measure changes in aggregate diversity (see Section~\ref{metrics_subsection}). First, we show the impact of MMDA re-ranking on each of the user and item sides by measuring various metrics. Secondly, we show the effectiveness of MMDA in dealing with user-item utility trade-offs. Also, we illustrate the impact of setting different capacity values for items.

\subsection{Datasets}

In our analysis we used three publicly available datasets. The first dataset is MovieLens 1M (ML)\footnote{https://grouplens.org/datasets/movielens}, a specific dataset for movie recommendation which is widely used in Recommender system's domain. ML contains 6,040 users and 3,702 movies and 1M ratings. Sparsity of ratings in this dataset is about 96\%.

Two other datasets used in our experiments are from Amazon categories \cite{he2016ups}: Amazon Books and Amazon Movies. The original Amazon books contains 2.5M users and about 1M items (books). Due to huge sparsity of this dataset we have randomly selected a subset of dataset in which each user has rated at least 20 items and each item has received at least 20 ratings. The resulting dataset reduced to 10K users and 9K items. The same process has been applied to the Amazon Movies dataset. Our sampled dataset contains 6.5K users and 6K items. The sparsity of the ratings are 72.3\% and 98.5\% for Amazon books and Amazon movies, respectively. 

\subsection{Experimental Settings}

MMDA is a post processing method which re-ranks the previously generated recommendations. We use two well-known collaborative filtering methods: User-based kNN \cite{sarwar2001item} and Non-negative Matrix Factorization (NMF) \cite{fevotte2011_NMF_algorithms, cichocki2009_NMF} as the underlying algorithms to generate the initial rankings. Utility of users are defined based on item rankings from these recommendations. On the other hand, utility of items can either be explicitly defined by item sellers, or they can be implicitly derived by recommending users to items \cite{vargas2014salesdiv}. Since, the seller preferences were not available to us, we used the implicit approach to define the item utilities.

For each dataset and recommendation method we first generated recommendations and then using the MMDA re-ranking method and predefined constraints (different $cap$s can be used for each item) we generated the final recommendations. The results are shown in total of 6 tables (Tables \ref{tab:MovieLensNMF} through \ref{tab:AmzBooksUserKNN}) corresponding to different dataset and recommender combinations.

We split the data into training, validation and testing sets with 80/10/10 percent ratios. For each dataset, we have tuned the parameters of base recommendation methods so that the accuracy (precision) is the maximized over the validation set. The parameters of the recommender methods used in our experiments are as follows. In MovieLens dataset we set the number of neighbors to 10 in User-based kNN. Also, in same dataset the number of latent factors is set to 40 when using NMF. For the Amazon datasets (both books and movies) we set the number of neighbors to 40 for kNN and the number of latent factors is set to 400 in NMF.

We used two baseline re-ranking methods to compare to our approach.

\textbf{Bayesian Re-ranking (BR)} \cite{vargas2014salesdiv}: Another re-ranking method specifically designed to increase catalog coverage. Given the score of each user-item pair $s(u,i)$, the BR re-scoring is defined by: 
\begin{equation}
    s_{BR}(u,i) = s(u,i) \Big( \sum_{u' \in \mathcal{U}}{s(u', i)} \Big)^{-\alpha}
\end{equation}
We have tuned the only hyper-parameter of this method $\alpha$ using grid search. Not all the values of $\alpha$ produce desired results. For example, setting values larger than $0.01$ does not change neither precision, nor gini-index. 

\textbf{Bi-criteria optimization using network-flow (NetFlow)} \cite{graphReRanking2017post}: We introduced this paper in the related work section. This method tries to optimize for two objectives: minimum discrepancy and maximum relevance and formulates the objective as $[\lambda \times discrepancy(H)$ $\minus \mu \times relevance(H)]$ (where $H$ stands for hyper-graph that is the original recommendations). We have tuned the two parameters ($\lambda$ and $\mu$) such that to obtain maximum accuracy at the various values of aggregate diversity (Gini-index in our experiments, see section~\ref{metrics_subsection}). This is exactly the same approach that we used for tuning the parameters of our method. The hyper-graph used in our experiments include all the recommended items available in the catalog and their corresponding scores obtained from the recommender system (either NMF or user-based kNN).

\begin{figure}[tbp]
    \centering
    \includegraphics[scale=0.27]{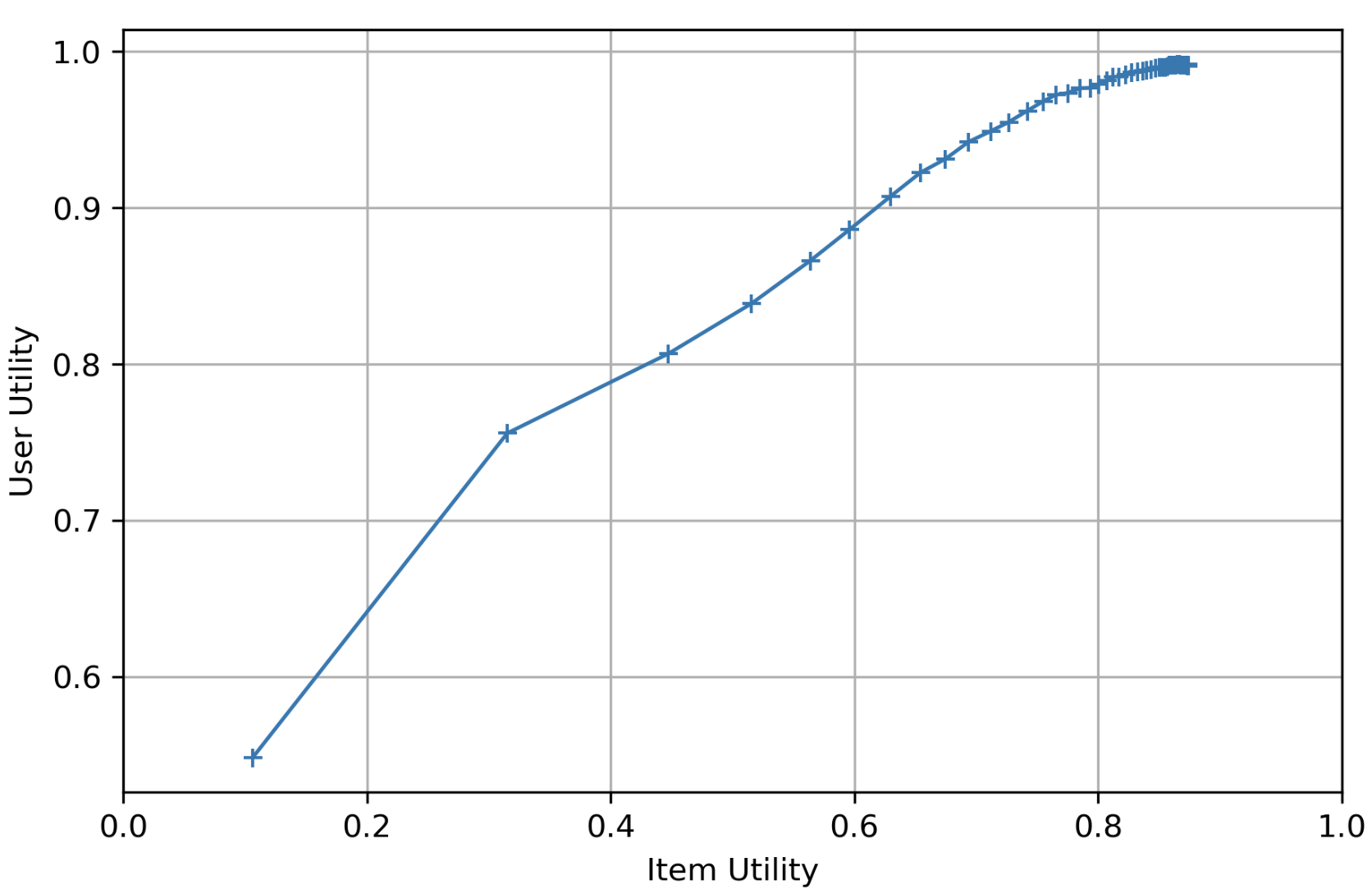}
    \caption{Amazon Movies: Utilities of both sides at each iteration of the MMDA, item cap=15, base recommender=NMF.}
    \label{fig: AmazonMVIterUtil}
\end{figure}

\begin{figure}
    \centering
    \includegraphics[scale=0.27]{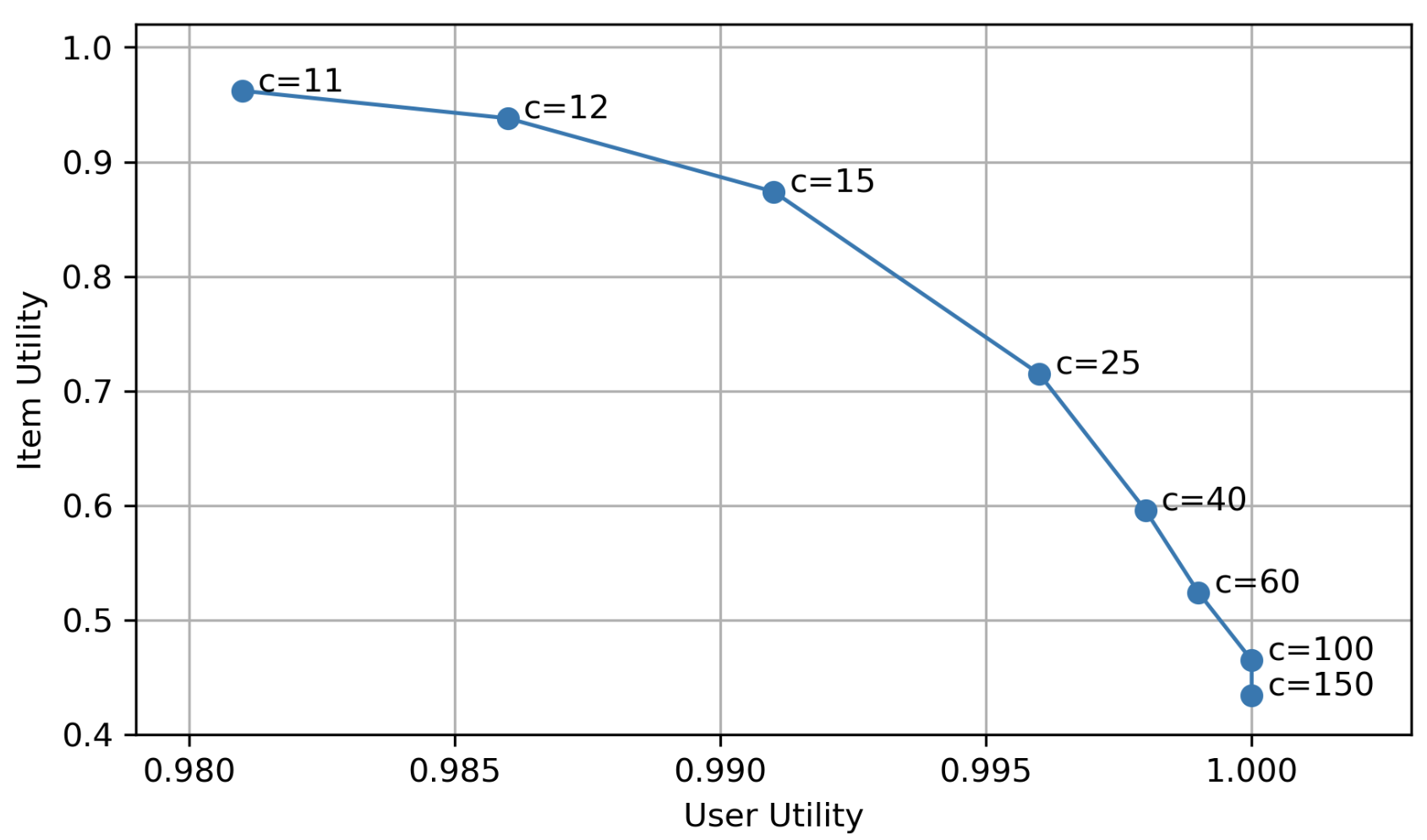}
    \caption{Amazon Movies: Utilities of both sides given various item caps c in MMDA.}
    \label{fig: AmazonMV_UtilVarCaps}
\end{figure}

\begin{figure}
    \centering
    \includegraphics[scale=0.28]{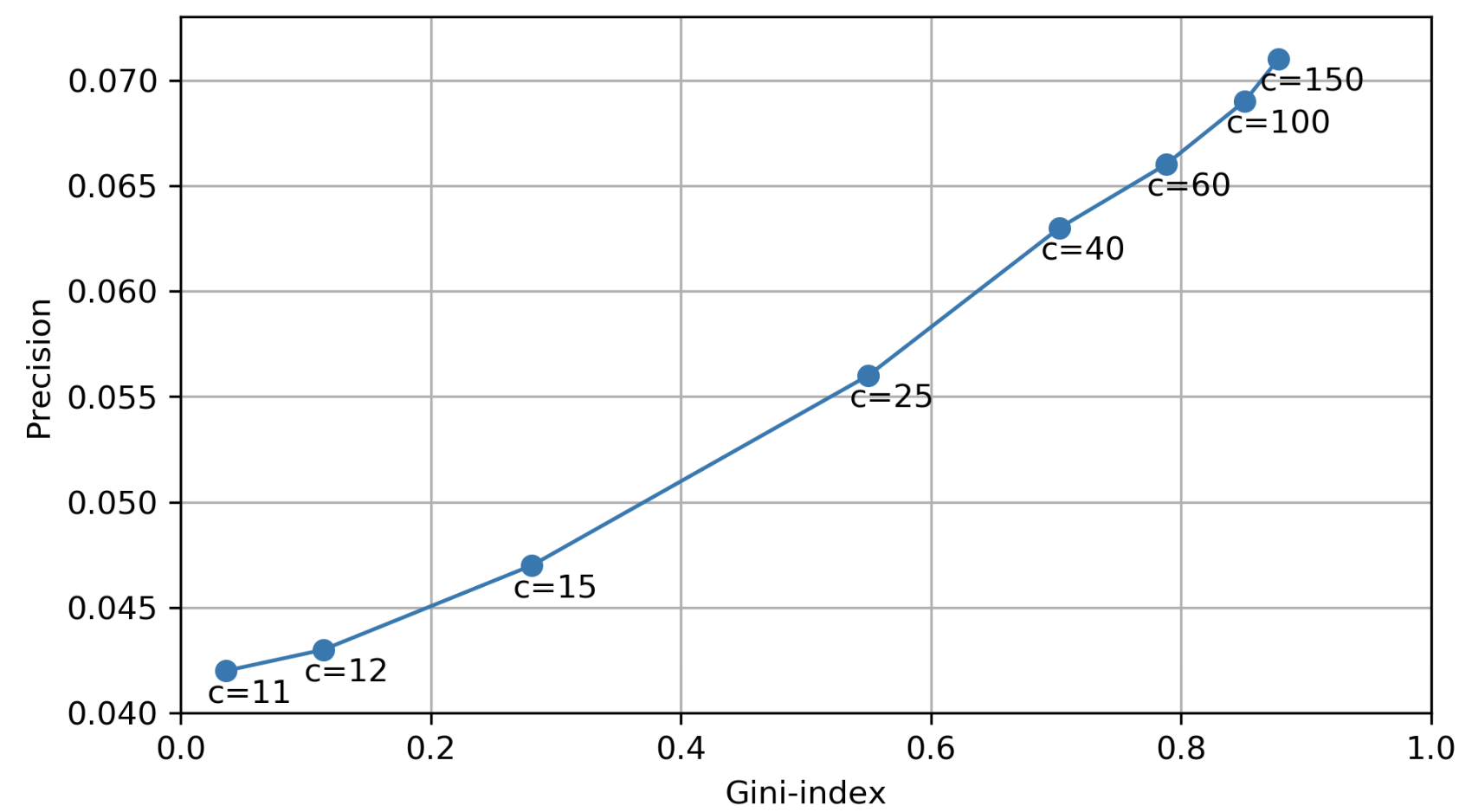}
    \caption{Amazon Movies: Precision@10 and Gini-index given various item caps c in MMDA.}
    \label{fig: AmazonMV_PrGiniVarCaps}
\end{figure}

\subsection{Evaluation Metrics} \label{metrics_subsection}

\noindent\textbf{Accuracy-based measures of utility:} \\

\noindent\textbf{User-side Utility ($Precision@k$)}: We use precision to measure the accuracy of re-ranked results. $Precision@k$ is the average of fraction of relevant items in each user's recommendation list $L_u$.
\begin{equation}
    Precision@k = \frac{1}{|\mathcal{U}|} \sum_{u \in \mathcal{U}}{|L_u \cap \mathcal{T}_u|}
\end{equation}
where $k = |L_u|$ is the recommendation list size and $\mathcal{T}_u$ is the user u's test set. 

\noindent\textbf{Aggregate Diversity Measures:} \\

\noindent\textbf{Gini-Index}: This is the measure of fair distribution of items into recommendation lists of all the users. The ideal (maximum fairness) case is when this distribution is uniform. The Gini-index of uniform distribution is equal to zero and so smaller values of Gini-index are desired.
\begin{equation}
    Gini = \frac{1}{|\mathcal{I}|} \sum_{k=1}^{|\mathcal{I}|}{(2k - |\mathcal{I}| - 1) \times p(i_k|L)},
\end{equation}
where $L$ is the set of all the recommendations for users and $p(i_k|L)$ is the probability of the $k$-th least recommended item being drawn from the recommendation lists in $L$:
\begin{equation}
   p(i|L) = \frac{|\{u \in \mathcal{U} | i \in L_u\}|}{\sum_{j \in \mathcal{I}}|\{u \in \mathcal{U} | j \in L_u\}|} 
\end{equation}

Catalog coverage is another widely used metric defined as the fraction of all the items that has been recommended at least once in all recommendation lists. Although, it used extensively as a measure of aggregate diversity, it is not a reliable measure. The main reason is that it is prone to noise. In a typical recommendation dataset, you would find thousands of users. Just by recommending an item once, it is not actually considered as covered. In that sense, Gini-index is much more reliable and more difficult metric to optimize. Note that in our tables in order to make results easier to understand the inverse of gini-index is shown. Therefore, the greater values of gini-index are more desired.   

\noindent\textbf{Intra-List Distance (ILD)}: This is the measure of individual user diversity \cite{ziegler2005improving, eskandanian2017clusteringDiv_evaluating}, which is a pairwise distance between all the items in each user's recommendation list.
\begin{equation}
    ILD(L_u) = \frac{1}{|L_u|(|L_u|-1)} \sum_{i \in L_u}\sum_{j \in L_u}{d(i,j)}
\end{equation}
Where $d(i,j)$ is the Cosine distance between two items based on ratings data. Note that, while our direct objective is not increasing individual diversity, we nevertheless measured ILD to explore possible correlations between aggregate diversity and individual diversity.

\noindent\textbf{User and Item Utilities:} We defined these utilities based on the notion of NDCG and the ranking distance from each side's ranking preferences. The precise definition was provided in Section~\ref{sec:definitions}.

\begin{table*}[] \small
  \centering
  \begin{tabular}{c | c c c c c c c c c c c c}
    & NMF & \begin{tabular}{@{}c@{}}BR \\ $\alpha$$=$$0.01$\end{tabular}& \begin{tabular}{@{}c@{}}BR \\ $\alpha$$=$$1e$$-$$3$\end{tabular}& \begin{tabular}{@{}c@{}}BR \\ $\alpha$$=$$4e$$-$$4$\end{tabular}&
    \begin{tabular}{@{}c@{}}NetFlow \\ $\lambda$$=$$2.0$ \\ $\mu$$=$$0.0$\end{tabular}  & \begin{tabular}{@{}c@{}}NetFlow \\ $\lambda$$=$$2.0$ \\ $\mu$$=$$0.01$\end{tabular} & \begin{tabular}{@{}c@{}}NetFlow \\ $\lambda$$=$$0.99$ \\ $\mu$$=$$0.01$\end{tabular} &
    \begin{tabular}{@{}c@{}}MMDA \\ $cap$$=$$17$\end{tabular}& \begin{tabular}{@{}c@{}}MMDA \\ $cap$$=$$20$\end{tabular}& \begin{tabular}{@{}c@{}}MMDA \\ $cap$$=$$50$\end{tabular} & \begin{tabular}{@{}c@{}}MMDA \\ $cap$$=$$80$\end{tabular} \\ [0.5ex]
    \hline
    Precision@10 &              22.4 & 6.5 & 20.6 & 25.2 & 2.2 & 16.7 & 21.7 & 8.7 & 9.7 & 14.7 & 17.9\\ 
    Gini(reversed) &            6.3 & 6.1 & 13.0 & 8.0 & 85.3 & 35.1 & 16.6 & \textbf{95.9} & 81.6 & 35.1 & 20.7\\
    $Avg(prec,gini)$ &        14.3 & 6.3 & 16.8 & 16.6 & 43.8 & 25.9 & 19.2 & \textbf{52.3} & 45.6 & 24.9 & 19.3\\
    \hline
    $Utility_\mathcal{U}$ &     100.0 & 97.9 & 99.8 & 99.9 & 74.3 & 98.7 & 99.7 & 85.0 & 89.4 & 99.2 & 99.3\\
    $Utility_\mathcal{I}$ &     16.9 & 14.5 & 28.3 & 23.4 & 65.1 & 56.0 & 38.0 & \textbf{84.4} & 73.8 & 47.9 & 24.0\\
    $Avg(Utilities)$ &              58.5 & 56.2 & 64.0 & 61.7 & 69.7 & 77.4 & 68.8 & \textbf{84.7} & 81.6 & 73.5 & 61.6\\
    \hline
    ILD &                   0.487 & 0.636 & 0.570 & 0.487 & 0.755 & 0.741 & 0.487 & \textbf{0.898} & 0.846 & 0.594 & 0.618\\
  \end{tabular}
  \caption{MovieLens 1M dataset, NMF recommendations.}
  \label{tab:MovieLensNMF}
\end{table*}

\begin{table*}[] \small
  \centering
  \vspace{-2em}
  \begin{tabular}{ c | c c c c c c c c c c c c}
    & kNN & \begin{tabular}{@{}c@{}}BR \\ $\alpha$$=$$0.01$\end{tabular}& \begin{tabular}{@{}c@{}}BR \\ $\alpha$$=$$1e$$-$$3$\end{tabular}& \begin{tabular}{@{}c@{}}BR \\ $\alpha$$=$$4e$$-$$4$\end{tabular}& 
    \begin{tabular}{@{}c@{}}NetFlow \\ $\lambda$$=$$2.0$ \\ $\mu$$=$$0.01$\end{tabular}  & \begin{tabular}{@{}c@{}}NetFlow \\ $\lambda$$=$$2.0$ \\ $\mu$$=$$0.01$\end{tabular} & \begin{tabular}{@{}c@{}}NetFlow \\ $\lambda$$=$$0.8$ \\ $\mu$$=$$0.2$\end{tabular} &
    \begin{tabular}{@{}c@{}}MMDA \\ $cap$$=$$17$\end{tabular}& \begin{tabular}{@{}c@{}}MMDA \\ $cap$$=$$20$\end{tabular}& \begin{tabular}{@{}c@{}}MMDA \\ $cap$$=$$50$\end{tabular} & \begin{tabular}{@{}c@{}}MMDA \\ $cap$$=$$80$\end{tabular} \\ [0.5ex]
    \hline
    
    Precision@10 &          22.4 & 3.1 & 17.5 & 21.6 & 2.5 & 6.7 & 9.1 & 9.4 & 10.1 & 14.6 & 16.9\\ 
    Gini(reversed) &        8.1 & 14.6 & 20.7 & 11.5 & 86.2 & 86.3 & 71.2 & \textbf{96.0} & 81.8 & 35.1 & 23.2\\
    $Avg(prec,gini)$ &      15.2 & 8.9 & 19.1 & 16.6 & 44.4 & 46.5 & 40.1 & \textbf{52.7} & 46.0 & 24.8 & 20.1\\
    \hline
    $Utility_\mathcal{U}$&  100.0 & 97.3 & 99.9 & 100.0 & 77.3 & 96.2 & 98.4 & 86.3 & 93.8 & 99.2 & 99.6\\
    $Utility_\mathcal{I}$&  31.2 & 30.7 & 51.1 & 43.2 & 66.0 & 74.1 & 71.6 & \textbf{82.8} & 73.3 & 47.9 & 41.1\\
    $Avg(Utilities)$ &          65.6 & 64.0 & 75.5 & 71.6 & 71.7 & \textbf{85.2} & 85.0 & 84.5 & 83.5 & 73.5 & 70.3\\
    \hline
    ILD &                   0.512 & 0.908 & 0.539 & 0.505 & 0.771 & 0.883 & 0.893 & \textbf{0.964} & 0.925 & 0.659 & 0.634\\
  \end{tabular}
  \caption{MovieLens 1M dataset, User-kNN recommendations.}
  \label{tab:MovieLensUserKNN}
\end{table*}

\begin{table*}[] \small
  \centering
  \vspace{-2em}
  \begin{tabular}{c | c c c c c c c c c c c c}
    & NMF & \begin{tabular}{@{}c@{}}BR \\ $\alpha$$=$$0.01$\end{tabular}& \begin{tabular}{@{}c@{}}BR \\ $\alpha$$=$$1e$$-$$3$\end{tabular}& \begin{tabular}{@{}c@{}}BR \\ $\alpha$$=$$4e$$-$$4$\end{tabular}& 
    \begin{tabular}{@{}c@{}}NetFlow \\ $\lambda$$=$$2.0$ \\ $\mu$$=$$0.01$\end{tabular}  & \begin{tabular}{@{}c@{}}NetFlow \\ $\lambda$$=$$0.8$ \\ $\mu$$=$$0.2$\end{tabular} & \begin{tabular}{@{}c@{}}NetFlow \\ $\lambda$$=$$0.01$ \\ $\mu$$=$$0.99$\end{tabular} &
    \begin{tabular}{@{}c@{}}MMDA \\ $cap$$=$$12$\end{tabular}& \begin{tabular}{@{}c@{}}MMDA \\ $cap$$=$$15$\end{tabular}& \begin{tabular}{@{}c@{}}MMDA \\ $cap$$=$$20$\end{tabular} & \begin{tabular}{@{}c@{}}MMDA \\ $cap$$=$$50$\end{tabular} \\ [0.5ex]
    \hline
    
    Precision@10 &          7.2 & 1.9 & 3.9 & 6.8 & 1.0 & 1.4 & 2.7 & 4.3 & 4.7 & 5.4 & 6.5\\ 
    Gini(reversed) &        11.1 & 15.3 & 32.0 & 20.4 & 85.7 & 85.5 & 7.5 & \textbf{88.6} & 71.9 & 55.2 & 24.6\\
    $Avg(prec,gini)$ &      9.1 & 8.6 & 17.9 & 13.6 & 43.3 & 43.5 & 5.1 & \textbf{46.4} & 38.3 & 30.3 & 15.5\\
    \hline
    $Utility_\mathcal{U}$&  100.0 & 99.0 & 99.6 & 99.9 & 85.7 & 87.3 & 98.2 & 98.6 & 99.1 & 99.5 & 99.9\\
    $Utility_\mathcal{I}$&  41.6 & 43.5 & 74.2 & 67.6 & 77.4 & 79.5 & 21.4 & \textbf{93.7} & 87.4 & 78.3 & 55.4\\
    $Avg(Utilities)$ &        70.8 & 71.2 & 86.9 & 83.8 & 81.6 & 83.4 & 59.8 & \textbf{96.2} & 93.2 & 88.9 & 77.7\\
    \hline
    ILD &                   0.530 & 0.783 & 0.666 & 0.562 & 0.946 & 0.962 & \textbf{0.965} & 0.819 & 0.818 & 0.767 & 0.707\\
  \end{tabular}
  \caption{Amazon Movies dataset, NMF recommendations.}
  \label{tab:AmzMoviesNMF}
\end{table*}

\begin{table*}[] \small
  \centering
  \vspace{-2em}
  \begin{tabular}{c | c c c c c c c c c c c c c}
    & kNN & \begin{tabular}{@{}c@{}}BR \\ $\alpha$$=$$0.01$\end{tabular}& \begin{tabular}{@{}c@{}}BR \\ $\alpha$$=$$1e$$-$$3$\end{tabular}& \begin{tabular}{@{}c@{}}BR \\ $\alpha$$=$$4e$$-$$4$\end{tabular} & \begin{tabular}{@{}c@{}}NetFlow \\ $\lambda$$=$$2.0$ \\ $\mu$$=$$0.0$\end{tabular}  & \begin{tabular}{@{}c@{}}NetFlow \\ $\lambda$$=$$0.99$ \\ $\mu$$=$$0.01$\end{tabular}  & \begin{tabular}{@{}c@{}}NetFlow \\ $\lambda$$=$$0.01$ \\ $\mu$$=$$0.99$\end{tabular} & \begin{tabular}{@{}c@{}}MMDA \\ $cap$$=$$12$\end{tabular}& \begin{tabular}{@{}c@{}}MMDA \\ $cap$$=$$15$\end{tabular}& \begin{tabular}{@{}c@{}}MMDA \\ $cap$$=$$20$\end{tabular} & \begin{tabular}{@{}c@{}}MMDA \\ $cap$$=$$50$\end{tabular} \\ [0.5ex]
    \hline
    
    Precision@10 &          7.1 & 1.4 & 4.9 & 6.9 & 1.1 & 3.5 & 7.0 & 5.4 & 5.8 & 6.2 & 7.0 \\
    Gini(reversed) &        13.9 & 21.4 & 51.5 & 28.7 & 85.4 & 85.8 & 14.1 & \textbf{88.9} & 73.6 & 58.5 & 30.1 \\
    $Avg(prec,gini)$ &    10.5 & 11.4 & 28.2 & 17.8 & 43.2 & 44.6 & 10.6 & \textbf{47.2} & 39.7 & 32.4 & 18.6 \\
    \hline
    $Utility_\mathcal{U}$&  100.0 & 98.7 & 99.7 & 99.9 & 88.6 & 99.4 & 100.0 & 99.2 & 99.5 & 99.6 & 99.9 \\
    $Utility_\mathcal{I}$&  49.0 & 30.0 & 77.2 & 71.1 & 63.2 & 74.5 & 49.5 & \textbf{81.4} & 76.6 & 72.2 & 60.5 \\
    $Avg(Utilities)$ &          74.5 & 64.3 & 88.5 & 85.5 & 75.9 & 87.0 & 74.8 & \textbf{90.3} & 88.0 & 85.9 & 80.2 \\
    \hline
    ILD &                   0.690 & 0.838 & 0.752 & 0.745 & 0.964 & \textbf{0.98} & 0.922 & 0.810 & 0.757 & 0.718 & 0.722 \\
  \end{tabular}
  \caption{Amazon Movies dataset, User-kNN recommendations.}
  \label{tab:AmzMoviesUserKNN}
\end{table*}

\begin{table*}[] \small
  \centering
  \vspace{-1em}
  \begin{tabular}{c | c c c c c c c c c c c c}
    & NMF & \begin{tabular}{@{}c@{}}BR \\ $\alpha$$=$$0.01$\end{tabular}& \begin{tabular}{@{}c@{}}BR \\ $\alpha$$=$$1e$$-$$3$\end{tabular}& \begin{tabular}{@{}c@{}}BR \\ $\alpha$$=$$4e$$-$$4$\end{tabular} & 
    \begin{tabular}{@{}c@{}}NetFlow \\ $\lambda$$=$$2.0$ \\ $\mu$$=$$0.0$\end{tabular}  & \begin{tabular}{@{}c@{}}NetFlow \\ $\lambda$$=$$2.0$ \\ $\mu$$=$$0.01$\end{tabular} & \begin{tabular}{@{}c@{}}NetFlow \\ $\lambda$$=$$0.8$ \\ $\mu$$=$$0.2$\end{tabular} &
    \begin{tabular}{@{}c@{}}MMDA \\ $cap$$=$$12$\end{tabular}& \begin{tabular}{@{}c@{}}MMDA \\ $cap$$=$$15$\end{tabular}& \begin{tabular}{@{}c@{}}MMDA \\ $cap$$=$$20$\end{tabular} & \begin{tabular}{@{}c@{}}MMDA \\ $cap$$=$$50$\end{tabular} \\ [0.5ex]
    \hline
    Precision@10 &          14.3 & 3.6 & 4.2 & 11.5 & 2.4 & 8.6 & 14.2 & 9.4 & 10.4 & 11.4 & 13.8\\ 
    Gini(reversed) &        18.2 & 16.1 & 20.3 & 33.7 & 87.5 & 85.8 & 19.9 & \textbf{96.2} & 77.5 & 59.0 & 26.6\\
    $Avg(prec,gini)$ &    16.2 & 9.9 & 12.2 & 22.6 & 45.0 & 47.2 & 17.0 & \textbf{52.8} & 44.0 & 35.2 & 20.2 \\
    \hline
    $Utility_\mathcal{U}$ & 100.0 & 99.4 & 99.5 & 99.9 & 90.9 & 99.6 & 100.0 & 98.6 & 99.5 & 99.8 & 99.9\\
    $Utility_\mathcal{I}$ & 47.8 & 37.8 & 61.9 & 74.7 & 85.6 & \textbf{97.3} & 52.1 & 96.3 & 88.2 & 77.7 & 53.9\\
    $Avg(Utilities)$ &        73.9 & 68.6 & 80.7 & 87.3 & 88.2 & \textbf{98.4} & 76.0 & 97.4 & 93.8 & 88.8 & 76.9 \\
    \hline
    ILD &                   0.845 & 0.921 & 0.921 & 0.897 & 0.97 & 0.939 & 0.845 & \textbf{0.94} & 0.906 & 0.903 & 0.883\\
  \end{tabular}
  \caption{Amazon Books dataset, NMF recommendations.}
  \label{tab:AmzBooksNMF}
\end{table*}

\begin{table*}[] \small
  \centering
  \vspace{-2em}
  \begin{tabular}{c | c c c c c c c c c c c c}
    & kNN & \begin{tabular}{@{}c@{}}BR \\ $\alpha$$=$$0.01$\end{tabular}& \begin{tabular}{@{}c@{}}BR \\ $\alpha$$=$$1e$$-$$3$\end{tabular}& \begin{tabular}{@{}c@{}}BR \\ $\alpha$$=$$4e$$-$$4$\end{tabular}& 
    \begin{tabular}{@{}c@{}}NetFlow \\ $\lambda$$=$$2.0$ \\ $\mu$$=$$0.01$\end{tabular}  & \begin{tabular}{@{}c@{}}NetFlow \\ $\lambda$$=$$0.99$ \\ $\mu$$=$$0.01$\end{tabular} & \begin{tabular}{@{}c@{}}NetFlow \\ $\lambda$$=$$0.8$ \\ $\mu$$=$$0.2$\end{tabular} &
    \begin{tabular}{@{}c@{}}MMDA \\ $cap$$=$$12$\end{tabular}& \begin{tabular}{@{}c@{}}MMDA \\ $cap$$=$$15$\end{tabular}& \begin{tabular}{@{}c@{}}MMDA \\ $cap$$=$$20$\end{tabular} & \begin{tabular}{@{}c@{}}MMDA \\ $cap$$=$$50$\end{tabular} \\ [0.5ex]
    \hline
    Precision@10 &          12.8 & 2.4 & 10.3 & 11.4 & 3.6 & 7.0 & 12.6 & 10.1 & 10.7 & 11.4 & 12.6 \\ 
    Gini(reversed) &        16.0 & 87.1 & 85.5 & 72.2 & 18.0 & 36.7 & 39.7 & \textbf{96.3} & 78.2 & 61.5 & 33.1\\
    $Avg(prec,gini)$ &    14.4 & 44.8 & 47.9 & 41.8 & 10.8 & 21.9 & 26.2 & \textbf{53.2} & 44.5 & 36.5 & 22.9 \\
    \hline
    $Utility_\mathcal{U}$ & 100 & 92.7 & 99.7 & 99.8 & 99.2 & 99.5 & 99.9 & 97.9  & 99.4 & 99.7 & 99.9\\
    $Utility_\mathcal{I}$ & 59.7 & 75.3 & \textbf{94.9} & 93.8 & 34.9 & 87.2 & 89.4 & 94.7  & 91.1 & 86.8 & 74.5\\
    $Avg(Utilities)$ &        79.8 & 84.0 & 97.3 & \textbf{96.8} & 67.0 & 93.3 & 94.7 & 96.3 & 95.2 & 93.2 & 87.2 \\
    \hline
    ILD &                   0.831 & 0.918 & 0.883 & 0.883 & 0.955 & 0.935 & 0.856 & \textbf{0.965} & 0.93 & 0.942 & 0.888\\
  \end{tabular}
  \caption{Amazon Books dataset, User-kNN recommendations.}
  \label{tab:AmzBooksUserKNN}
\end{table*}

\subsection{Results and Discussion}

Figure \ref{fig: AmazonMVIterUtil} shows the increase in user and item utilities over successive iterations of MMDA. Interestingly, in the beginning both utilities are increasing at the same pace but over time the slope of the lines shows more gain in item utility versus user utility. This is perfectly aligned with our observations \ref{Obs_usersGetWorse} and \ref{Obs_itemsGetBetter} in the previous section.

Also, we ran experiments on the Amazon Movies data set to observe the effect of specifying various constraints for item capacity values $cap$ in MMDA. Note that since this is an assignment problem (meaning that for each user there are $k$ empty positions to assign items into those positions as recommendations) the $cap$ value cannot be smaller (lower bound) than $|\mathcal{U}| \times k / |\mathcal{I}|$ where $k$ is the recommendation list size. Choosing values smaller than the lower bound will result in empty recommendation lists for some users. Figures \ref{fig: AmazonMV_UtilVarCaps} and \ref{fig: AmazonMV_PrGiniVarCaps} show these results in terms of utilities (item-side versus user-side) and accuracy-diversity trade-off, respectively. In figure \ref{fig: AmazonMV_PrGiniVarCaps} the range of Gini-index values in x-axis are much larger than precision values in y-axis. This shows the effectiveness of MMDA in distributing items such that the loss in accuracy is minimized. To the best of our knowledge no other re-ranking method is capable of competing with MMDA when very low Gini values (close to uniform distribution of items) is demanded. This would be the major advantage of the MMDA compared to the baseline methods as shown in the tables.

Figure \ref{fig: AmazonMV_UtilVarCaps} shows the performance of MMDA given various $cap$s. Note that the range of change in user utility is much smaller compared to the range of changes in item utility. This reflects the effectiveness of MMDA in managing the trade-off in utilities such that with little loss in user-side utility, the gain in item-side utility would be substantial.

Among all of the existing methods in the literature our method is the only one that can approach uniform distribution of items (close to zero on the Gini-index in figure \ref{fig: AmazonMV_PrGiniVarCaps}). More interestingly, this huge lift in aggregate diversity does not result in a huge loss of accuracy. Therefore, our approach effectively deals with the accuracy-diversity trade off. Also, the increase in aggregated diversity increases individual recommendation lists' diversity which is beneficial both to users and item providers.

Results on tables 1-6 show a comparative evaluation of different re-ranking methods using metrics discussed in section \ref{metrics_subsection}. Note that in order to make the tables more understandable we reversed the values of gini-index such that larger values of any metric are desired. Tables \ref{tab:MovieLensNMF} and \ref{tab:MovieLensUserKNN} show the results for MovieLens dataset. The concentration bias of ratings over popular items in training set of this dataset is much more than other two datasets. This makes managing the trade-off between accuracy and diversity more challenging. We can see all of the baseline methods are bound from below in terms of Gini values. The Gini values of MMDA on the other hand are not restricted to a specific range. For example, the Gini values for MMDA with $cap=17$ and $cap=20$ are much smaller representing near uniform item distribution across recommendations with the cost of losing about 12\% in precision. (also compare it to the baselines) On the other datasets, the concentration bias is slightly smaller and therefore MMDA is able to maintain a high level of precision while decreasing Gini values (thus increasing aggregate diversity) significantly. 

Even though our goal was not to increase individual diversity, ILD measures in almost all of our results show the effectiveness of MMDA in increasing Intra-list diversity of recommendations. This suggest that both user and item sides are benefiting from MMDA re-ranking results.


\section{Conclusion and Future Works}

We have proposed a two-sided re-ranking recommendation approach that takes into account both item and user utilities in order to produce the best possible match between items and users. Based on an extension of the Deferred Acceptance algorithm for solving the well-known stable matching problem, our method is guaranteed to always produce a stable match in an efficient manner. We have proved this property of the algorithm and we have also shown that it is user-optimal subject to specified item cap constraints. These properties allow our algorithm to automatically manage the trade-off between aggregate diversity objective on the item side and the accuracy objective on the user side. This is a post-processing approach that can be used in conjunction with any underlying collaborative recommendation algorithm, adding to its flexibility. Our experimental evaluation using three datasets has provided additional evidence for the effectiveness of our approach in increasing aggregate diversity (and more generally item utilities) while improving or maintaining recommendation accuracy.

In future work, we intend to experiment with additional base recommendation algorithms to determine the impact of those algorithms on item and user utilities resulting from the matching algorithm. In addition we will explore the possibility of using different underlying recommendation algorithms on the item and user sides to improve the accuracy of initial item and user preference orderings, respectively.

\balance
\bibliographystyle{ACM-Reference-Format}
\bibliography{ref}

\end{document}